\documentclass[conference]{IEEEtran}
\IEEEoverridecommandlockouts
\usepackage{cite}
\usepackage{amsmath,amssymb,amsfonts}
\usepackage{algorithmic}
\usepackage{graphicx}
\usepackage{textcomp}
\usepackage{xcolor}

\usepackage{nameref}

\usepackage{soul}

\newcommand\noteeh[1]{{\textcolor{black}{#1}}}

\def\vs{\emph{vs. }}

\newcommand{\pb}[1]{\vspace{0.75ex}\noindent{\bf \em #1}\hspace*{.3em}}

\usepackage[hyphens]{url}

\usepackage{todonotes}
\usepackage{tabularx}
\usepackage{booktabs}
\usepackage{subcaption}
\usepackage{caption}
\usepackage{xcolor}
\usepackage{color,soul}

\usepackage{color, colortbl}
\definecolor{Gray}{gray}{0.9}

\usepackage{color}
\definecolor{lightgray}{gray}{0.75}
\usepackage{amsthm}

\def\BibTeX{{\rm B\kern-.05em{\sc i\kern-.025em b}\kern-.08em
    T\kern-.1667em\lower.7ex\hbox{E}\kern-.125emX}}

\usepackage{fancyhdr}
\usepackage{kantlipsum}
\fancypagestyle{plain}{
\fancyhf{}
\fancyhead[C]{Conference on \LaTeX} %% C or L or R.
}
\usepackage{eso-pic}

\begin{document}

\AddToShipoutPictureBG*{
\AtPageUpperLeft{
\setlength\unitlength{1in}
\hspace*{\dimexpr0.5\paperwidth\relax}%% change \dimexpr0.5\paperwidth\relax appropriately
\makebox(0,-0.75)[c]{\textbf{2022 IEEE/ACM International Conference on Advances in Social
Networks Analysis and Mining (ASONAM)}}}}

\title{
% Psychologists, Therapists, Writers, Doctors? 
Exploring Mental Health Communications among Instagram Coaches }

% \and
% \and
% \and
% \and
% \and

\author{
\IEEEauthorblockN{
Ehsan-Ul Haq\IEEEauthorrefmark{1}, 
Lik-Hang Lee\IEEEauthorrefmark{2}, 
Gareth Tyson\IEEEauthorrefmark{4}, 
Reza Hadi Mogavi\IEEEauthorrefmark{1}, 
Tristan Braud\IEEEauthorrefmark{1}, 
and Pan Hui\IEEEauthorrefmark{1}\IEEEauthorrefmark{3}\IEEEauthorrefmark{4}}
\IEEEauthorblockA{\IEEEauthorrefmark{1}Hong Kong University of Science and Technology, HKSAR}
\IEEEauthorblockA{\IEEEauthorrefmark{2}Korea Advanced Institute of Science and Technology}
\IEEEauthorblockA{\IEEEauthorrefmark{3}University of Helsinki, Helsinki}
\IEEEauthorblockA{\IEEEauthorrefmark{4}Hong Kong University of Science and Technology, Guangzhou}
Email: \{euhaq,rhadimogavi\}@connect.ust.hk  \quad 
likhang.lee@kaist.ac.kr \quad \{gtyson,braudt,panhui\}@ust.hk
}

\maketitle
\IEEEoverridecommandlockouts
\IEEEpubid{\parbox{\columnwidth}{\vspace{8pt}
\makebox[\columnwidth][t]{IEEE/ACM ASONAM 2022, November 10-13, 2022}
\makebox[\columnwidth][t]{978-1-6654-5661-6/22/\$31.00~\copyright\space2022 IEEE} \hfill}
\hspace{\columnsep}\makebox[\columnwidth]{}}
\IEEEpubidadjcol

% \IEEEoverridecommandlockouts
% \IEEEpubid{\parbox{\columnwidth}{\vspace{8pt}
% \makebox[\columnwidth][t]{IEEE/ACM ASONAM 2022, November 10-13, 2022}
% \makebox[\columnwidth][t]{\url{http://dx.doi.org/10.1145/XXXXXXX.XXXXXXX}}
% \makebox[\columnwidth][t]{978-1-6654-5661-6/22/\$31.00~\copyright\space2022 IEEE} \hfill}
% \hspace{\columnsep}\makebox[\columnwidth]{}}
% \IEEEpubidadjcol

% \IEEEoverridecommandlockouts
% \IEEEpubid{\parbox{\columnwidth}{\vspace{8pt}
% \makebox[\columnwidth][t]{IEEE/ACM ASONAM 2022, November 10-13, 2022}
% \makebox[\columnwidth][t]{\url{http://dx.doi.org/10.1145/XXXXXXX.XXXXXXX}}
% \makebox[\columnwidth][t]{978-1-6654-5661-6/22/\$31.00~\copyright\space2022 IEEE} \hfill
% \hspace{\columnsep}\makebox[\columnwidth]}}
% \IEEEpubidadjcol

\begin{abstract}
There has been a significant expansion in the use of online social networks (OSNs) to support people experiencing mental health issues. This paper studies the role of Instagram influencers who specialize in coaching people with mental health issues. Using a dataset of 97k posts, we characterize such users' linguistic and behavioural features. We explore how these observations impact audience engagement (as measured by likes). We show that the support provided by these accounts varies based on their self-declared professional identities. For instance, Instagram accounts that declare themselves as \textit{Authors} offer less support than accounts that label themselves as a \textit{Coach}. We show that increasing information support in general communication positively affects user engagement. However, the effect of vocabulary on engagement is not consistent across the Instagram account types. Our findings shed light on this understudied topic and guide how mental health practitioners can improve outreach. 
\end{abstract}


\section{Introduction} \label{sec:introduction}

The US government reports that one in five Americans are facing a mental health problem~\cite{mythsfacts}. Similar statistics have been reported in other regions~\cite{uk_survey_health_2014}.
Mental health issues can have devastating consequences, yet such issues are often considered taboo.\footnote{\url{https://www.swissre.com/risk-knowledge/risk-perspectives-blog/world-mental-health-day-preventing-breaking-taboos.html}}
This can result in a reluctance to seek help from professionals, especially since some societies do not encourage discussions on mental health issues. In such cases, the use of information and communication technologies (ICT) 
can be an important asset in helping people. Researchers have studied Online Social Networks' (OSNs) role in the support and prediction of mental health issues~\cite{santarossa2018lancerhealth,de2014characterizing}. Social media has also led to the emergence of mental health `influencers', i.e., OSN profiles that focus on offering mental health advice.
Such accounts frequently post advice and interact with their audience in an attempt to promote well-being.

Despite this, there is a paucity of (quantified) knowledge about how these users operate and impact others. Considering the critical role these accounts can play in people's lives, we argue that it is vital to study them.
We take two social science theories as motivating factors for our research. First, we look at these accounts from the \textit{identity} perspective. Professional identity helps individuals define and express their role as a professional and social entity~\cite{colbeck2008professional,caza2016construction}. Moreover, individuals with professional roles often enjoy trust from the wider society~\cite{larson1979rise}. This leads us to reason that while the goal of these accounts is to promote mental health, their professional identity can be used to characterize such accounts. Thus, taking this self-declared \textit{identity} as our comparative scale to study their communications.

Second, we look at the psychological characteristics of the language used. The language used for mental health support has an effect on patient health~\cite{miller2017analyzing,valdes2010analysis}. \noteeh{The analysis of language related to mental health is widely carried out using the psychological characterization of vocabulary and quantifying the level of support~\cite{de2017language}, more particularly in the context of \textit{Information and Emotion Support} that can have practical outcomes~\cite{peng2020exploring}. In our work, we analyze the communication based on the level of \textit{Information and Emotion Support} provided by mental health influencers.} We note that while these online communications largely remain one-way in contrast to an actual therapy session, the primary goal of health influencers is to promote well-being. Hence, the role of language remains important.

This paper presents the first analysis of how mental health influencers use Instagram. We choose \emph{Instagram} because of the growing anecdotal evidence of the importance it plays in this field.\footnote{\url{https://lifegoalsmag.com/10-best-mental-health-and-therapy-accounts-on-instagram}} In addition to the characterization of the language, and support provided by these accounts, we further quantify the effect of these attributes on user engagement. 
Specifically, we ask two research questions: 
\begin{itemize}

    \item \textit{\noteeh{RQ1: Is there any difference in the language across accounts based on their \textit{identity}? More particularly, what are the levels of information and emotional support provided by these accounts?}}
    % {RQ1: What kind of content (from a psychological perspective) is discussed in posts by health professionals? ~\noteeh{More particularly, is there any difference in the content across the accounts?} } (\S\ref{subsec:linguistics}) %Specifically, we look into linguistic characterization, and quantify the level of informational and emotional support provided in these posts.

    \item \textit{RQ2:~\noteeh{What is the effect of language, \textit{identity}, and support provided by these accounts on user engagement?}} 
    % \gareth{Is this RQ what you really mean? It looks broken, with Identity in the middle? together, added the comma}
    
    % \item \textit{RQ2: What are the linguistic and account-specific factors that impact the level of user engagement?} (\S\ref{subsec:rq2})
    
\end{itemize}

To answer these questions, we gathered three years of data, covering a curated set of 127 Instagram accounts of practising mental health professionals. We explore their activities across several axes, focusing on the linguistic characteristics of post captions used by accounts. First, we label all posts using  LIWC semantic categories~\cite{pennebaker2001linguistic} for linguistic analysis and support levels~\cite{peng2020exploring}. We then measure the presence of these semantic categories in posts and explore how they vary across influencers. For example, we find that mental health professionals tend to use high levels of emotive language (e.g., love) and affiliation terminology (e.g., friends). This leads us to explore such behavioural traits' impact on audience engagement. To this end, we perform regression across various features to understand the impact on engagement. 
The results reveal that audiences engage more with content that has Cognitive semantics. Critically, we find that merely providing informational and emotional support is \emph{not} enough and that different account types (e.g., Therapist \vs  Writers) experience different reactions to particular semantic categories of posts.
We argue that these results can inform future guidelines about how to compose a message to increase audience engagement and reach a larger audience.

%A deeper dive in to the analysis shows that different semantic categories at different accounts have varied engagement results. The account personification and the post timings affect the user-engagement with such content. 

% younger generations and demographics. 

% the role of online forums to promote and  to compare the strategies employed by the IG account holders.

\section{Background \& Related Work} \label{sec:related_work}

\pb{Background.} 
The use of information technologies in health support~\cite{Kostoska2018real} and diagnosis~\cite{kao2020user} is widely studied. In recent years, mental health and digital wellbeing have become a prominent research area~\cite{de2013predicting}, including the development of various digital support systems~\cite{cavlo2016computing,doherty2008technologies,alqahtani2019usability}. A significant part of this has been the growing use of OSNs to offer support and insights to different organisations~\cite{haq_2022_govt}. It has been shown that platforms like Instagram provide a space for users to express themselves~\cite{villani_does_2012}, and help overcome societal barriers~\cite{doherty2008technologies}.

Several works have studied the (positive) effects of health support~\cite{krause2001social,kaplan1977social,glanz2008health}. 
To measure this, researchers have developed metrics of social support~\cite{nick2018online,peng2020exploring,harandi2017correlation}.
For example, Nick et al. categorize online social support into four categories: emotional support, companionship support, informational support, and instrumental support~\cite{nick2018online}. Glanz et al. categorize the behaviour as emotional, instrumental, informational, and appraisal ~\cite{glanz2008health}.
Borrowing from these, in this paper, we explore two major categories of support: 1) Informational Support and 2) Emotional Support~\cite{peng2020exploring,harandi2017correlation}.

\pb{Related Work.}
There have been numerous studies of Instagram. These cover, for example, studying social activism \cite{kandappu2018feasibility,haq_screenshots_2022}, tourism \cite{rossi2018venice}, linguistic differences \cite{zhang2019language}, influencers~\cite{zarei2020characterising} and predicting the personality traits of users \cite{ferwerda2015predicting}. 
%There exist %have been 
Moreover, several studies particularly focus on %looking at 
the promotion of (self-)body image satisfaction on Instagram~\cite{baker2019qualitative,ahadzadeh2017self,santos2020instagram}. 
For instance, \cite{baker2019qualitative} conducts a thematic analysis to understand female users' beauty standard(s) on their self-posted images. Similarly, another work explores gender inequalities on Instagram~\cite{santos2020instagram}. In contrast, this paper focuses on the use of Instagram by health care influencers, and further probes the user engagements of their followers.

Others have looked at Instagram's roles in promoting health matters and well-being of the public~\cite{pinto2020instagram,griffith2021mentalhealthart,pinto2021public,andalibi2017sensitive}. 
For example, Griffith et al. analyze the effect of visual features on promoting health by analysing the engagement on posts~\cite{griffith2021mentalhealthart}. 
They show that both visual and textual (caption) features are used to promote mental health related messages. The use of Instagram to promote well-being within a community has further been shown to help in improving health-related behaviour~\cite{santarossa2018lancerhealth}, largely because of homophilic tendencies~\cite{centola2011experimental}. For instance, Andalibi et al. find that Instagrammers show a sense of community support towards posts related to depression~\cite{andalibi2017sensitive}, where Instagram acts as a user-accepted and low-barrier enabler of disclosing one's sensitive or emotional issues. 

\begin{figure}[t]
    \centering
    \includegraphics[width=.48\textwidth]{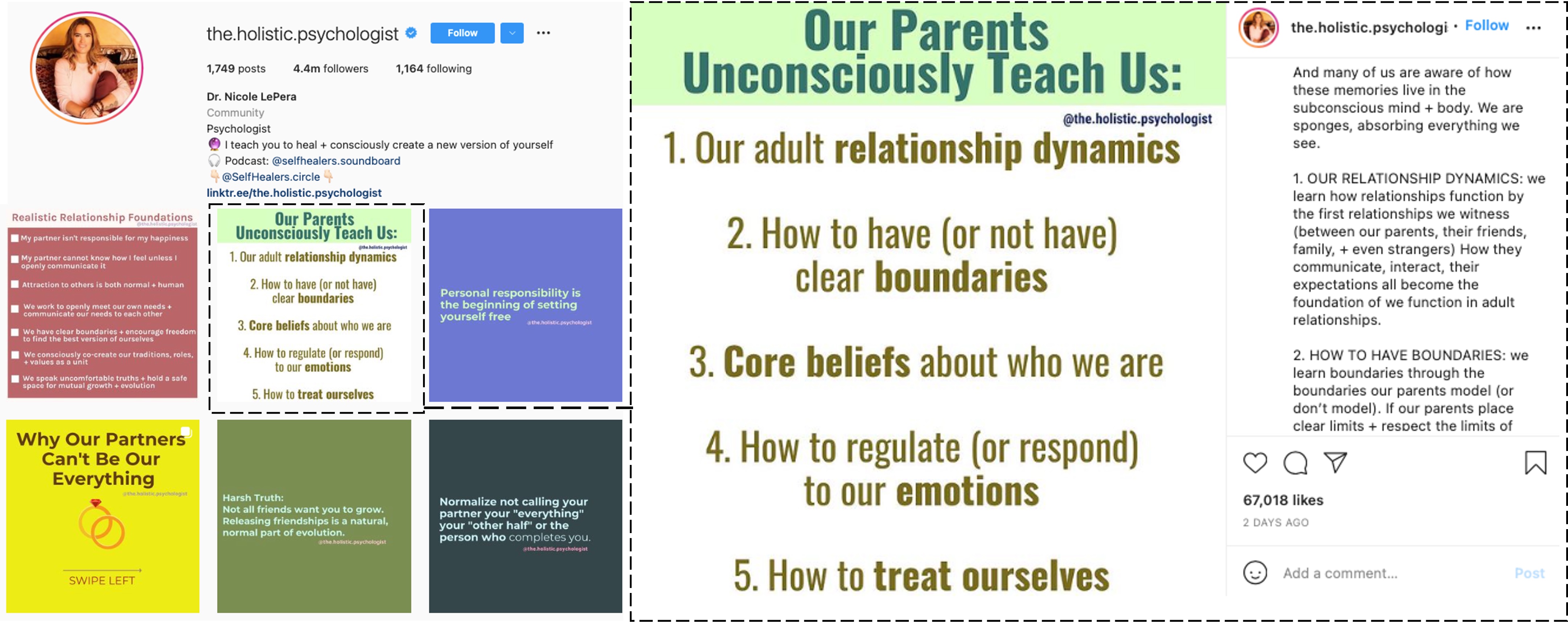}
    \caption{(Left) The emerging style of mental health `influencers' on an Instagram account named \textit{the.holistic.psychologist}. (Right) An example of pixelised texts (i.e., texts embedded in an image) and its captions (repeated) in the post.}
    \label{fig:exexex}
    
\end{figure}

Our research builds on this prior work, with a focus on evaluating the activities of mental health influencers on Instagram. In contrast to these prior works that primarily analyse the efficacy of image-driven content sharing, we focus on the linguistic features of \emph{captions} written by a wide range of accounts (e.g., \textit{Psychologists}, \textit{Therapists}, \textit{Writers}, and \textit{Doctors}, see Figure~\ref{fig:exexex} for reference). %We specifically employ the LIWC dictionary and the theory of informational/ emotional support to explore the messages inside \textbf{captions}.
%It is important to note that mental health influencers primarily \textbf{embedded summarised texts in images, repeated in the captions}. Hence, our work uniquely contributes to the understanding of this emerging style of content delivery, primarily as pixelised texts in the images (for an example, see Figure~\ref{fig:exexex}). %account account  ...
% \gareth{Can you add some more specifics in here? Prior text was quite vague. }

%Our work particularly stands out at as it evaluates the qualified experts communications on a platform that is often described with negative outcomes. Our empirical analysis characterizes the content and linguistic characteristics that are well-received by their audience, such findings can serve as guidelines for new practitioners and other automated service providers. 

% In addition, The presence of government health agencies on such platform may contain the political messages as in the case for Protugese and Brazil~\cite{pinto2021public}. 

% some studies have also explored it to predict the users' mental health. In addition to such predictive analysis, Instagram has been used to 

% Apart from negative behavior associations, Instagram  about cyberbullying~\cite{gupta2020temporal}.

\section{Dataset \& Annotations}
\label{sec:dataset}

\subsection{Data Collection}
We start by searching Instagram for the \textit{mentalheatlh} hashtag. From these results, we collect a list of all verified accounts that have at least 10K followers (as we are interested in influencers) and use English language for communication. We aim to select accounts that are from mental health professionals.
Thus, we inspect the self-description on the profiles to check for relevant titles such as Dr., PhD, Therapist etc. If the self-description contains any external links to websites or other online presence for the account holder, we further check that link. 
We take a rigorous approach, whereby we only include accounts that we are highly confident are genuine professionals.  
Following this, we use snowball sampling to find more accounts.
Specifically, we inspect other accounts that these seed users follow or tag in their posts, resulting in a total of 127 professional accounts.

We then retrieve all posts from these 127 accounts (in the span between the 1st January 2019 and November 2021, with data collected in November 2021) using the CrowdTangle API.\footnote{\url{https://www.crowdtangle.com/}}
In total, we retrieve 97,918 posts from all 127 accounts, consisting of 9,116 videos, 75,269 photos, 13,533 albums (may include photos and videos together), and 2,423 IGTV posts. %Albums may include photos and videos together. 
% Note, the dataset includes all the default information provided by the Crowdtanlge API.
% (this includes the caption, number of likes/comments, and number of followers of the page at the time of a post). 

% \pb{Ethical Consideration:} The profile data in this paper is publicly available. However, we understand that these posts may contain comments or other links that can lead to sensitive and identifiable information concerning an individual's privacy and health. 
% Thus, we only inspect influencer posts (rather than comments or responses). 
% We use data according to CrowdTangle's privacy policy and do not share the data to prevent increased privacy risk.

%In our paper, we utilise the the textual description of the posts. 
%We pre-process the text to exclude hashtags and emojis from the text. The mean word-count in textual description is 136 with median word count of 108, and maximum of 450 words.

\subsection{Annotations}
\label{sec:method}

\emph{Professional Identity:} Our exploratory analysis shows that account holders use various titles in their self-description, e.g., \textit{Therapist, Coach, Writer, or Doctor}. We first perform a manual analysis of the account self-description to extract the title used. However, mere qualitative extraction of these titles does not result in any specific categorization, as a large percentage of accounts use more than one title in their description. To make it more systematic, we train word embeddings based on complete account descriptions using FastText~\cite{bojanowski2016enriching}. We then measure the word vectors for the manually extracted labels for each account; we take the average vector in case of more than one label. For instance, if an account holder describes themself as \textit{Doctor and coach}, we take word-vectors for each of these and take an average. We use the elbow method to find the optimum number of clusters for K-Means clustering (clusters = 5) to group similar accounts. In the rest of the analysis, we refer to these clusters as users with type 1, 2, and so on. Examples of the accounts into these clusters are shown in Table~\ref{tab:profile_clusters}. We note that `authors' may not initially appear to be related to our study, however, recall that we manually curated the list of accounts to ensure mental health support function.

\begin{table}[t]
    \centering
    \scriptsize
    \begin{tabular}{l|l|l}
    \toprule
        \textbf{Type} & \textbf{Accounts}&\textbf{Titles used} \\
        \midrule
        Type 1 &  11 &  Author \\
        Type 2 &  47 &  Combination of Dr. with Therapists and Psychologists\\
        Type 3 &  39 &  Combination of Therapist, Psychologist, Coach \\
        Type 4 &  11 &  Coach \\
        Type 5 &  19 &  Therapist \\
         
         \bottomrule
    \end{tabular}
    \caption{Users' profile classification into five categories based on self-description.}
    \label{tab:profile_clusters}
\end{table}

In addition, we count the referral links to other social media platforms in the Instagram self-description and the presence of links to professional websites. In our dataset, 123 accounts contain a link to a personal website. We note that most accounts in our dataset (60\%) do not use links to other platforms in their Instagram profiles, and 28\% of accounts do not use links to other platforms on their website.

\section{Methodology \& Results}
\label{sec:results}

We first analyse the linguistic characteristics of posts (RQ1), before building a series of regression models to better understand the factors that impact audience engagement (RQ2).

\subsection{RQ1: Content Characterization} \label{subsec:linguistics}

We first explore what kind of content is discussed in posts by health professionals.

\pb{Semantic Attributes.} We analyze the linguistic artifacts of Instagram posts by assigning semantic categories to the most frequently used words in the dataset.
These categories highlight the psychological themes of the posts. We pre-process all posts by removing hashtags, mentions, punctuation, emojis, and URLs. 
We then associate each remaining word with its LIWC
semantic category~\cite{pennebaker2015development}. Note that we do not distinguish between LIWC categories and subcategories, as main category is not exclusive to set of sub-category words. To normalize the effect of highly active users, we look at the unigrams according to their uniformity of usage among all users
by Shanon's Entropy~\cite{zhang2019language}. This entropy measures the uniformity of a word among all the users. A higher value for a word means that the word is used uniformly among all users, while a lower entropy means fewer users use that word (Entropy distribution: mean = 0.8, std. deviation=1.19, min=0, max=4.8, 75\%=1.32). We do not include the word belonging to the article and preposition categories. This serves as a check that the linguistic feature and analysis represent the dataset. Words and their entropy values are shown in Table~\ref{tab:entropy_values}.

\begin{table}[t]
\scriptsize
\begin{tabular}{ll|ll|ll}
\toprule
\textbf{Word} & \textbf{Entropy} & \textbf{Word} & \textbf{Entropy} & \textbf{Word} & \textbf{Entropy} \\
\midrule
you & 4.78 & yourself & 4.68 & because & 4.69 \\
your & 4.76 & people & 4.66 & link & 4.23 \\
can & 4.81 & who & 4.75 & being & 4.77 \\
our & 4.80 & them & 4.74 & bio & 4.18 \\
what & 4.78 & time & 4.80 & these & 4.73 \\
not & 4.78 & life & 4.67 & way & 4.80 \\
have & 4.80 & self & 4.72 & body & 4.28 \\
when & 4.78 & one & 4.81 & dont & 2.43 \\
how & 4.79 & their & 4.71 & things & 4.70 \\
more & 4.56 & know & 4.78 & work & 4.74 \\
they & 4.73 & get & 4.78 & make & 4.79 \\
all & 4.82 & need & 4.74 & may & 4.63 \\
but & 4.76 & was & 4.67 & relationship & 4.43 \\
feel & 4.73 & there & 4.77 & other & 4.77 \\
its & 4.55 & just & 4.77 & help & 4.71 \\
will & 4.76 & some & 4.78 & see & 4.77 \\
love & 4.65 & want & 4.73 &  &  \\
\bottomrule
\end{tabular}
\captionof{table}{Top 50 uniformly used words across the dataset.}
\label{tab:entropy_values}
\end{table}

We observe a range of intuitive words that commonly occur, including things like `self', `know', `relationships' and `help'.
Table~\ref{tab:all_categories} summarises the most common categories observed. 8\% of words (unigrams) fall into the Affective Processes category. This covers both positive and negative statements related to emotion-laden behaviour. The second most common category is Drives (6.9\% of words); this category covers words that are associated with affiliations, rewards or risks, e.g., words like `friends', `success', and `bully'. The third most common category, covering 5.2\% words, is Relativity (relativ) which contains words related to time, space or motion, such as `arrive', and `season'. The fourth-ranked category (5.1\% of words) is Cognitive Process (cogproc). This consists of words related to reasoning, insights and certainty, e.g., `think', `because', and `always'. Biological Processes (bio) category covers words related to the body and health. The Social Processes (social) category includes words related to family, friends or gender-based references. 
A Kruskal-Wallis test confirms a significant difference in the distribution of these top-10 LIWC categories \( (\chi^2(9, N=97918) = 413166, p< 10^{-16} )\).

We note that all of the top-10 LIWC semantic categories fall under ``Psychological process'' in LIWC. This shows that these communications are primarily focused on mental health-related topics (such as `anxiety', and `trauma'). The group also contains words that show the causation and healing process, with frequent mentions of social and biological processes. We observe words like `us', `people', and `one' appearing frequently. These tend to show that the communication of influencers is not directed towards individuals but at the wider community. 
We also see words like `bio' and `links'. `Bio' refers to the biography of the Instagram account and contains links to authors' websites. These have options for bookings, online therapies, podcasts, etc. The self-descriptions 
show that while the influencers try to share messages related to mental health, they also seek to share and promote their resources. 

\begin{figure}[t]
     \centering
     \begin{subfigure}[b]{0.15\textwidth}
         \centering
         \includegraphics[width=\textwidth]{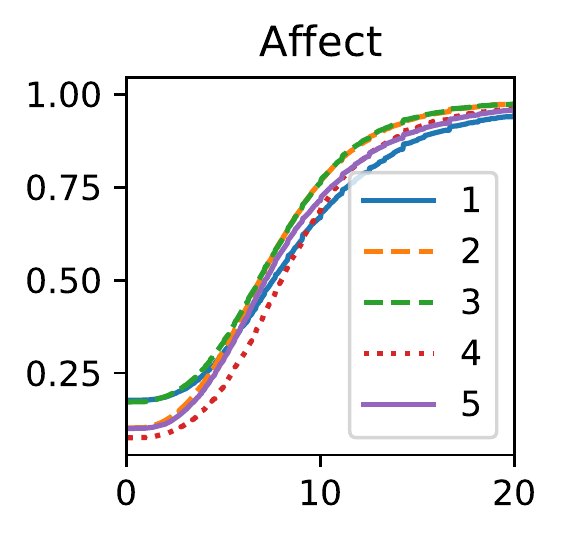}
         \label{fig:affect}
     \end{subfigure}
    %  \hfill
     \begin{subfigure}[b]{0.15\textwidth}
         \centering
         \includegraphics[width=\textwidth]{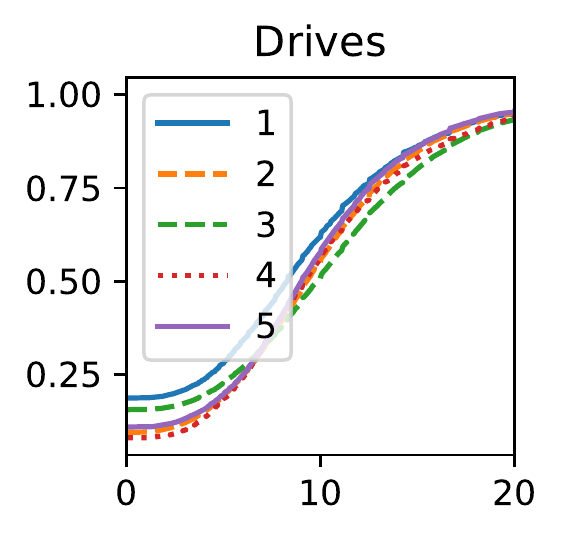}
         \label{fig:drives}
     \end{subfigure}
    %  \hfill
     \begin{subfigure}[b]{0.15\textwidth}
         \centering
         \includegraphics[width=\textwidth]{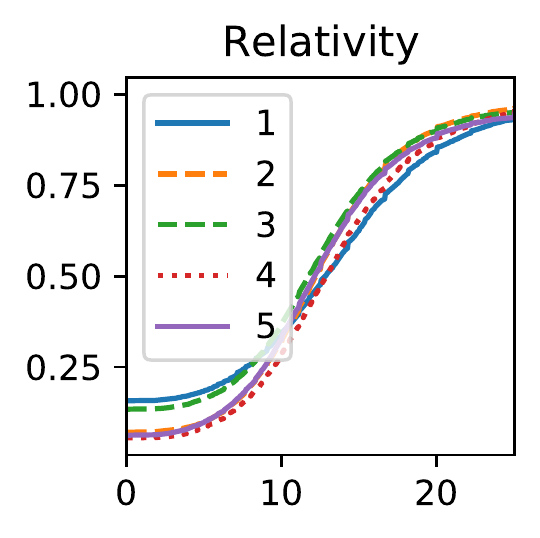}
         \label{fig:relativ}
     \end{subfigure}
        \caption{CDFs for word distribution for Top-10 LIWC categories in each of 5 account types.}
        \label{fig:cdfs_clusters}
\end{figure}

\pb{Account Specific Content.} We next use the LIWC dictionary to analyse the content difference across the accounts. We use the Kruskal-Wallis test followed by Dunn's posthoc (adjusted with the Bonferroni method to control the multiple comparisons) analysis to study the difference across the account types. The posthoc analysis for pair-wise comparison is reported in Table~\ref{tab:dunn_test_statistics}, along with the Cumulative Distribution Function (CDFs) of data distribution in Figure~\ref{fig:cdfs_clusters} for the top three categories (Affective Processes, Drives, and Relativity). 
The Kruskal-Wallis test shows that the use of specific vocabulary is significantly different across the account. The pair-wise comparisons, along with the distribution analysis for the top-10 categories, show that number of words in a given sentence related to \textit{Biological Process}, \textit{Negative Emotions} and \textit{Work} is higher in \textit{Type-5} accounts. The use of \textit{Power} related words is higher in posts from \textit{Type-2} accounts. \textit{Drives} is used more often in posts from \textit{Type-3} accounts. The rest of the LIWC categories (\textit{Affect, Relativity, Social, Cognitive Processing} and \textit{Positive Emotions}) in a given sentence are usually higher in \textit{Type-4} account categories. 

This result suggests that the \textit{Type-4} and \textit{Type-5} accounts usually have a broader range of communication compared to other account types. The higher use of words related to \textit{Cognitive Process} relates to reasoning and insights, whereas the \textit{Drives} focuses on the achievements and risks. For instance, a post from \textit{Type-3} accounts, having at least five words related to \textit{Power} category, discusses the challenges of facing a dilemma of choices, e.g., \textit{``It is normal to feel two or more things at once. The struggle to understand that can be challenging- the “both and” scenario . We can practice feeling two things. Joyful about eating a bagel and scared about eating a dead food. Nervous about gaining weight and excited to be following a meal plan appropriately. Scared to enter recovery and relieved to be receiving help.''}

\begin{table}[t]
\scriptsize
\centering
\begin{tabular}{l|l|l}
\toprule
\textbf{Category} & \textbf{Count} & \textbf{Percentage} \\
\midrule
Affective Process (affect) & 3305 & 8.5 \\
Drives (drives) & 2700 & 6.9 \\
Relativity (relativ) & 2017 & 5.2 \\
Cognitive Process (cogproc) & 1994 & 5.1 \\
Biological Process (bio) & 1885 & 4.8 \\
Negative Emotions (negemo) & 1873 & 4.8 \\
Social Process (social) & 1482 & 3.8 \\
Positive Emotions (posemo) & 1389 & 3.6 \\
Work (work) & 1295 & 3.3 \\
Power (power) & 1143 & 2.9\\
\bottomrule
\end{tabular}
    \caption{10 most frequent LIWC categories in complete dataset. Percentages are based on whole dataset with some words belonging to more than one category. }
    \label{tab:all_categories}
\end{table}

\begin{table*}[t]
\scriptsize
\centering
\begin{tabular}{lllllllllllllllllllll}
\toprule
\textbf{Group} & \textbf{Affect}  & \textbf{Drives}  & \textbf{Relativ}  & \textbf{Cogproc}  & \textbf{Bio}  & \textbf{negemo}  & \textbf{social}  & \textbf{posemo}  & \textbf{work}  & \textbf{power}  \\
\midrule
\(\chi^2(4)\) & 809.22	*** &	448.66	***	&466.34	***&	430.75	***	&1485.3	***&	1263.1	***	&338.46	***&	1185.1	***	&766.2	***&	301.6	***\\
\midrule
1-2 & 10.30 *** & -16.79  *** & 4.39  *** & -2.38   & -1.83   & -25.53  *** & 0.57   & 19.80  *** & -13.61  *** & -16.76  *** \\
1-3 & 14.66  *** & -20.49  *** & 12.97  *** & -7.22  *** & 11.60  *** & -21.34  *** & 1.39   & 23.93  *** & 1.22   & -13.34  *** \\
2-3 & 6.55  *** & -5.72  *** & 12.51  *** & -7.06  *** & 19.40  *** & 5.51  *** & 1.21   & 6.43  *** & 21.18  *** & 4.58  *** \\
1-4 & -8.78  *** & -14.61  *** & -5.33  *** & -17.19  *** & -11.14  *** & -17.90  *** & -8.42  *** & -2.69   & -0.13   & -13.13  *** \\
2-4 & -21.19  *** & -1.33   & -11.01  *** & -18.95  *** & -11.99  *** & 3.31  * & -11.01  *** & -23.13  *** & 13.45  *** & 0.47   \\
3-4 & -25.39  *** & 2.64   & -19.48  *** & -13.79  *** & -25.20  *** & -0.54   & -11.68  *** & -27.21  *** & -1.38   & -2.69 &  \\
1-5 & 1.95   & -11.34  *** & 3.24  * & -8.09  *** & -16.93  *** & -35.07  *** & 9.01  *** & 18.75  *** & -14.94  *** & -13.98  *** \\
2-5 & -9.40  *** & 4.60  *** & -0.84   & -7.87  *** & -20.13  *** & -16.44  *** & 11.19  *** & 1.64   & -3.83  ** & 1.10   \\
3-5 & -14.43  *** & 9.08  *** & -10.82  *** & -2.07   & -35.21  *** & -20.50  *** & 10.00  *** & -3.52  * & -20.66  *** & -2.58   \\
4-5 & 11.58  *** & 4.68  *** & 9.09   & 10.77  *** & -4.71  *** & -15.43  *** & 18.24  *** & 21.69  *** & -14.80  *** & 0.42  \\
\bottomrule
\end{tabular}
\captionof{table}{Kruskal and Dunn's Test statistics for categories across different account types, \(( * = p < 0.01\), \(** = p < 0.001\), p-values \(*** < 0.0001)\)}
\label{tab:dunn_test_statistics}
\end{table*}

% From the accounts perspective, \textit{Type-1}. \textit{Type-2}. \textit{Type-3}. \textit{Type-4}. \textit{Type-5}

The most used categories are in line with the analysis of real therapy sessions~\cite{valdes2010analysis}. This suggests that communications from such accounts on social media contain linguistic cues that are suitable for people with mental health support requirements.

\pb{Support Provided by Professionals.} 
We next seek to measure the level of support present in the posts. Following previous work~\cite{peng2020exploring}, we study two kinds of support: Information (IS) and Emotion Support (ES). 
We thus use the classifier from Peng et al.~\cite{peng2020exploring} to label our posts into three levels (low, medium, and high) for both information and emotional support. This classification model is built using the 64 features extracted by the LIWC 2015 library. Two separate models are used: Random Forest for information labelling and XGboost for emotion labelling. We note that the original method is trained on using LIWC features, as LIWC is a dictionary-based method, and LIWC categories are assigned to the unigrams. Hence, the general difference in the platform or community language will be minimum when using the same model for Instagram posts, as the words are considered in individual scope rather than in a context. Additionally, a word category will remain consistent.
Note, information support refers to the providing of \textit{advice, referrals or knowledge}, whereas Emotion support covers discussion related to \textit{understanding, encouragement, affirmation, sympathy, or caring}~\cite{wang2012stay,peng2020exploring}.

\begin{figure}[t]
    \centering
    \includegraphics[width=.35\textwidth]{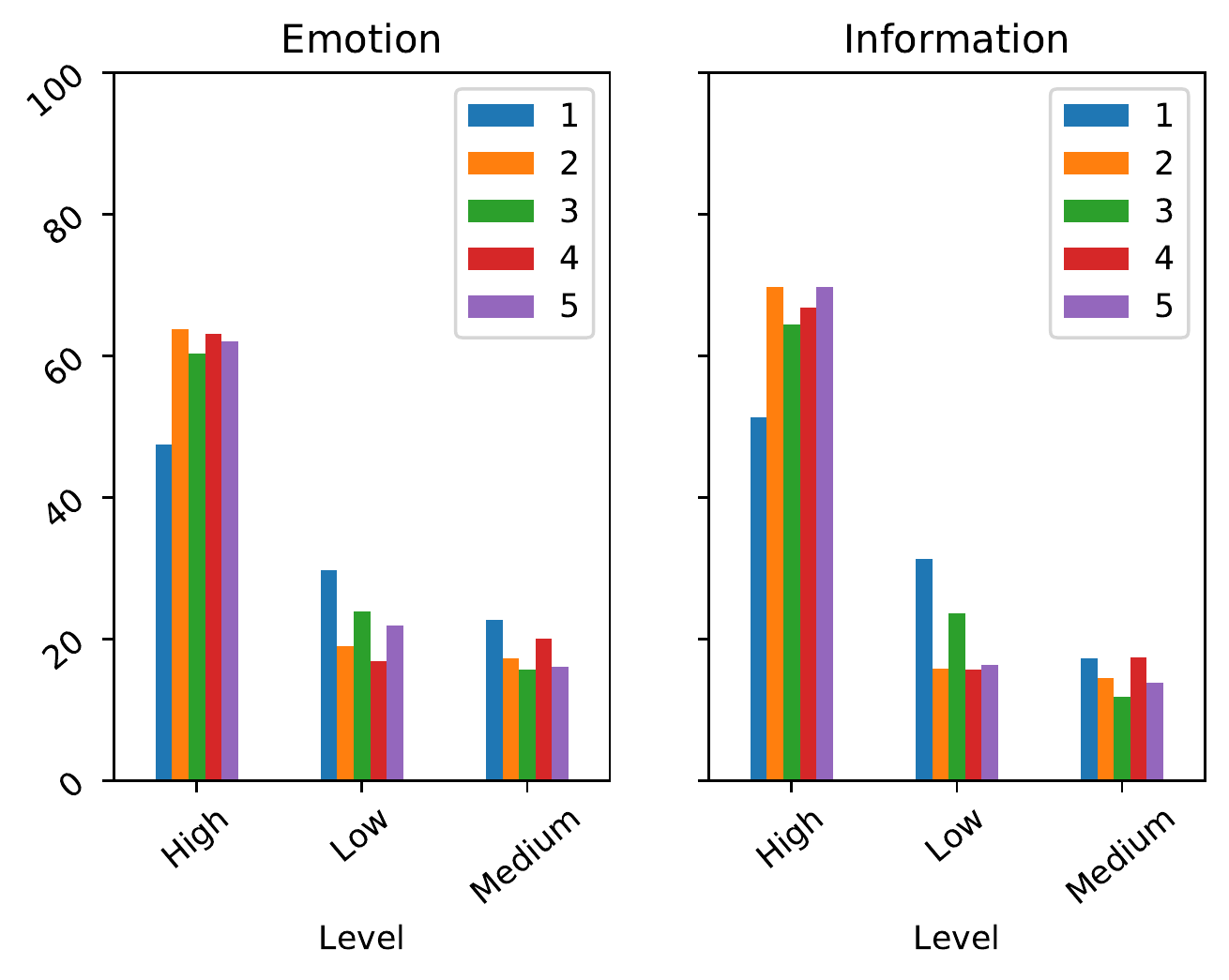}
    \caption{Label Distribution across the different account types (Type-1 to Type5), shown as percentage (y-axis) of total posts: most of the posts contain high levels of support.}
    \label{fig:support_distribution}
\end{figure}

The distribution of emotional and informational support across the whole dataset is shown in Figure~\ref{fig:support_distribution}. We see that the majority of posts are classified as containing both high ES and IS. We also show the distribution of such messages across our account types: a chi-square test shows the distribution across the account for support level is significantly different (Emotion= \(\chi^2 = 200, p<10^{-5}\), Information= \(\chi^2 = 82.8, p<10^{-5}\)). 
We find that for \textit{Type-1} have a difference of at least 10\% in sharing high ES and IS as compared to other accounts.

The use of features such as pronouns, social references, and positive emotions significantly affect the level of IS and ES~\cite{peng2020exploring}. The higher use of pronouns is associated with lower IS but higher ES. Social and Positive Emotions significantly affect IS and ES, respectively. When we look at these features and the results of LIWC-categories usage across the account types, we see the \textit{Type-1}%type-0 
accounts have a lower number of words per sentence in all of the categories; hence, in general, resulting in lower ES and IS. In contrast, the remaining accounts have high levels of \emph{both} ES and IS in their communications (at least 50\% of posts have a high level for IS and ES). We assume this might be because professionals' primary identity and communication focus, and expertise vary. %In general, the percentage of posts with high IS is larger than posts with high ES across the account categories. This can be explained by the higher use of positive emotion words across the accounts.
For instance, the following post from \textit{Type-4} accounts, having at least five positive emotion words, has a high ES and low IS: \textit{``Smile so hard your eyes close. Love so hard it makes you forget your past heartaches. If that’s what happens, why does it matter who creates this for you? It doesn’t! Love is love! Happy Pride Month!''}

\subsection{RQ2: Factors Driving the Engagement}
\label{subsec:rq2}

\pb{Methodology.}
We rely on the number of likes as a measure of engagement.
We formulate our problem as a regression task with the number of likes as the dependent variable (DV) and various post features as the independent variables (IV). 
A regression model estimates the change in the value of a dependent variable (either positive or negative) depending on the change in the value of an independent variable.
Here, we strive to understand how determinant emotional, account and support features are in driving the level of engagement. We normalize the number of likes based on the number of followers of the account at the time the post was created (to avoid bias/overfitting for highly popular accounts). 
As the number of likes in our dataset does not follow the Gaussian distribution, we use the negative binomial regression model that can take into effect this dispersion in the data and is commonly used for social media engagement modeling~\cite{bakhshi2014faces}. 
We then train several regression models with different feature sets to better understand which behaviors impact engagement. Note that the LIWC features in our models represent the number of words appearing in a message belonging to a given category. Hence, a positive/negative regression result should be associated with a one-word (or unit) change in the category that is being examined.

\begin{table}[t]
\scriptsize
\centering
\begin{tabular}{l|ll|r}
\toprule
\textbf{Variable} & \textbf{\(\beta\)} &  & \textbf{std.error} \\
\midrule
(Intercept) & 1.449 & *** & 0.012 \\
affect & -0.014 & *** & 0.004 \\
drives & 0.000 &  & 0.001 \\
relativ & -0.004 & *** & 0.000 \\
cogproc & -0.002 & *** & 0.000 \\
bio & 0.004 & *** & 0.001 \\
negemo & 0.011 & ** & 0.004 \\
social & -0.003 & *** & 0.001 \\
posemo & 0.004 &  & 0.004 \\
work & -0.012 & *** & 0.001 \\
power & 0.001 &  & 0.001 \\
\midrule
Deviance & \multicolumn{3}{r}{114091 }    \\
Residual Deviance & \multicolumn{3}{r}{113551 } \\
Log-likelihood & \multicolumn{3}{r}{-467304.2} \\
\bottomrule
\end{tabular}
\caption{Engagement for linguistic categories. (\(** = p < 0.01\), \(*** = p < 0.001\)).}
\label{tab:engagement_top_10_categories}
\end{table}

\pb{Semantic Category Effect.}\label{subsec:linguistic_engagement}
We start by training a model using 
several LIWC categories to represent each post (as IVs). A generalized model for predicting the number of likes can be described as follows:

\[ 
    log(Likes) = \theta + \sum_{i \in \mathbf{C_{}} }{[\beta_i {C_{i}} ]} + \epsilon 
\]

\noindent where \(C_{}\) represents the set of variables used in the model. Here, \(C_{}\) contains 10 variables representing the top-10 categories from the Table~\ref{tab:all_categories}. \(\theta\) and \(\epsilon\) represent the intercept and error terms for the model. For each post in the dataset, we use the LIWC dictionary to count the number of words under each category. Thus, for each post, we have an input vector with 10 features. We then train our first negative binomial regression model with the number of likes as the dependent variable.

The results are presented in Table~\ref{tab:engagement_top_10_categories}.
The model fits well, and the coefficients confirm that some categories do have an effect on the engagement of a post. The $\beta$ columns show the likely impact in the number of likes (increase or decrease) with one unit change in the respective variable.
Biological Processes, Emotions (positive and negatives) attract higher engagement.
Out of the top 10 LIWC categories: Work, Social Processes, Relative Processes, Affective Processes, and Cognitive Processes have a negative effect on the number of likes. The most positive effect is associated with negative emotions (negemo).
Further, Power and positive emotions (posemo) do not have a statistically significant impact on engagement levels. 
Our observations indicate that audiences engage more with content that is directly focused on health and emotions. Adding more topics (e.g., social or work) reduces engagement. Similarly, the addition of one word related to Negative Emotion increase the engagement 0.01x. The addition of a word related to the `Biological Process' can increase engagement by 0.004x. One level increase in the words related to Affective Process can decrease the engagement by 0.01x.

Note that Negative Emotion words do not necessarily convey a negative sentiment. This LIWC category includes vocabulary such as `anxiety', `trauma', and `anger'. We assume that being more explicit on these topics increases the interest of the audience and hence the number of likes. To highlight this, we randomly select a post from the dataset that has at least 5 Negative Emotion words, shown in Table~\ref{tab:category_example} (top). We highlight the negative words in yellow, showing that the overall message is largely informational.

\begin{table}[t]
    \centering
    \scriptsize
    \begin{tabular}{p{8cm}}
    \toprule
        \rowcolor{Gray}
         \textbf{Message with Negative Emotion category highlighted} \\
                  \hline
\textit{"\hl{Anxiety} is a major public health \hl{problem} that is reaching epidemic levels in the United States. The \hl{loss} to our society from these illnesses is staggering in terms of individual \hl{pain}, family strife, school and relationship failure, \hl{lost} work productivity, and death. \hl{Anxiety} is a Brain Illness . Our work and the research of many others has demonstrated that \hl{anxiety} is a brain illness, not the result of a \hl{weak} will or character problem. In addition to the common symptoms listed in the questions above, \hl{anxiety} can cause \hl{irrational} \hl{fears} or \hl{phobias} that become a burden. People with “pure \hl{anxiety}” tend to avoid anything that makes them \hl{anxious} or \hl{uncomfortable}, such as places or people that might trigger \hl{panic} attacks or interpersonal conflict. People with this type tend to predict the \hl{worst} and look to the future with \hl{fear}. They may be excessively \hl{shy} or \hl{startle} easily, or they may freeze in emotionally charged situations"} \\

        \bottomrule

    \end{tabular}
    \caption{Examples of messages with highlighted LIWC Negative Emotion and Social categories.}
    \label{tab:category_example}
\end{table}

\pb{Account Category Effect.} 
We next seek to understand the impact of the account category on engagement, i.e., our five categories based on the self-declared identities of the accounts. 
To include the account category type in the model, we introduce it as an interaction term in our model from the previous subsection. As such, we compare the effect of each semantic category within the account category. A generalised model, in this case, can be described as:

\[ 
    log(Likes) = \theta + \sum_{i \in \mathbf{C_{-}} }\sum_{j \in \mathbf{A_{-}} }{[\beta_i {C_{i}}{A_{j}} ]} + \epsilon 
\]

\noindent 
where \(A_{}\) represents the set of five Instagram account categories. To compare the engagement effect across different account categories, Type-1 is used as a baseline category, and all other account categories are compared against it. This means that a positive (negative) effect in any regression coefficient $\beta_{i}$ for any other category $A_{i}, i \in {\{Type-1, Type-2, .. , Type-5\}}$ will imply the possible increase (decrease) in engagement for $A_{i}$ compared to the Type-1 category for the same parameter(s). Note that we use numerical categories for each account type and the regression model automatically selects the baseline category in ascending order. Hence, the Type-1 category is selected as the baseline. 

First, we see that the account category does have a significant effect on the engagement level. Overall, variation exists in terms of categories getting higher engagement across the account types. Regression results shows that \textit{Affective Process} is likely to increase engagement for \textit{Type-5} accounts by 0.02x time as compared to \textit{Type-1}.~\footnote{All variables can be seen online. \url{https://tinyurl.com/rq2-variables}} \textit{Drives} is likely to increase the engagement for \textit{Type-4} by 0.03x, but likely to reduce the engagement for \textit{Type-3} accounts by 0.02x. \textit{Negative Emotions} likely increase the engagement for \textit{Type-3 and 4} by 0.02x and 0.1x as compared to \textit{Type-1}.

To summarise the effects of engagement across the accounts, we note that Type-2 get higher engagement for `Social Process', and `Power'. For Type-3, the categories with the highest positive effect on engagement are 'Biological Process`, and `Positive Emotions'. For Type-4, `Negative Emotions', `Drives', and `Work' result in higher engagement as compared to other accounts. For Type-5, 'Affective Process, Relativity and Cognitive Process' are likely to result in higher engagement. We also note that for these accounts, the higher engagement is not always in their most used semantic category. For instance, the usage of \textit{Cognitive Process} is greater for \textit{Type-4} accounts; however, in terms of engagement \textit{Cognitive Process} negatively affects the likelihood of higher engagement for these accounts.

\begin{table}[]
\centering
\scriptsize
\begin{tabular}{l|ll|l}
\toprule
\textbf{Variable} & \textbf{\(\beta\)} & \textbf{} & \textbf{S.E.} \\
\midrule
(Intercept) & 1.34 & *** & 0.03 \\
IS & -0.01 &  & 0.02 \\
ES & 0.01 &  & 0.02 \\
Type-2 & 0.12 & ** & 0.04 \\
Type-3 & -0.36 & *** & 0.04 \\
Type-4 & -0.20 & *** & 0.05 \\
Type-5 & 0.12 & * & 0.05 \\
\hline
IS:Type-2 & 0.02 &  & 0.02 \\
IS:Type-3 & 0.19 & *** & 0.02 \\
IS:Type-4 & 0.07 & * & 0.03 \\
IS:Type-5 & -0.09 & *** & 0.03 \\
ES:Type-2 & -0.10 & *** & 0.02 \\
ES:Type-3 & -0.04 &  & 0.02 \\
ES:Type-4 & -0.12 & *** & 0.03 \\
ES:Type-5 & -0.07 & ** & 0.03 \\
\bottomrule
Null Deviance &  \multicolumn{3}{r}{115181}   \\
Residual Deviance  &  \multicolumn{3}{r}{113313}   \\
2log-likelihood & \multicolumn{3}{r}{-465991.01} \\
\bottomrule
\end{tabular}
\caption{Emotion Support (ES) and Information Support (IS) interaction together with account category. Increase in information is positively associated with the engagement. (\(*=p < 0.05\), \(** = p < 0.01\), \(*** = p < 0.001\))}
\label{tab:support_level_prediction}
\end{table}

\pb{Information and Emotion Support Effect.}
We next look at the effect of Information and Emotion Support levels provided by the Instagram accounts on engagement received on posts. As a feature set, we take the three labels (high, medium, low) for both types of support, computed using~\cite{peng2020exploring} (see Section~\ref{subsec:linguistics}). A Kruskal-Wallis test on the normalized engagement shows a significant difference across the three ES levels (high, medium, and low) = (\(\chi^2(2;N= 97,918) = 65.714, p <0.0001)\) and across IS levels (high, medium, and low)= (\(\chi^2(2;N= 97,918) = 117.75, p <0.0001)\). 

This confirms that the number of likes received on these posts does vary. Thus, we expand our previous regression model to measure the effect of emotion/information support (together with the account categorization type) on the engagement level. To this end, we create a feature for each post consisting of its label for support level (high, medium, and low) together with the category of the account. 

Table~\ref{tab:support_level_prediction} shows the results of this model. When IS and ES are looked at together with the account characteristics, it has a varying effect on engagement. For instance, an increase in Information Support by one level (e.g, low to medium) increases the engagement for \textit{Type-3} accounts by 0.2x, as compared to \textit{Type-1} accounts. However, for \textit{Type-5} accounts, an increase in IS will likely decrease engagement. Interestingly, four account types have negative predictors as compared to \textit{Type-1} account types. Suggesting the increase in emotional support has a negative impact as compared to information support. Regardless, \textit{Type-3} accounts are least likely to be susceptible to negative impact as compared to the other accounts. This also suggests that audiences tend to prefer informational content over emotional support.

We repeat our regression analysis and report the results in Table~\ref{tab:profile_features}. The model reports statistically significant results, with both features having a negative effect on engagement. The results show that a 1x unit increase in the number of social media links in the account description reduces engagement by almost 0.3x. Similarly, if the number of links to other social media increases on the external websites of authors, the engagement on Instagram decreases 0.14x. 
The exact reason for this is unclear. We conjecture that followers may move to other platforms (via these links), rather than continue to engage on Instagram, thereby lowering the engagement on Instagram.

\begin{table}[]
\scriptsize
\centering
\begin{tabular}{l|ll|l}
\toprule
 \textbf{Variable} & \(\beta\) &  & \textbf{S.E.} \\
\midrule
Intercept & 1.04 & *** & 0.02 \\
links in self-description & -0.26 & *** & 0.02 \\
links on web & -0.14 & *** & 0.02 \\
\hline
% Null Deviance& \multicolumn{3}{r}{56380} \\
% Residual Dev, & \multicolumn{3}{r}{51130} \\
2Log-likelihood & \multicolumn{3}{r}{-281920} \\
\bottomrule
\end{tabular}
\caption{Engagement estimate based on social media links on profile and on the personal website. Engagement is negatively associated with both features. (\(*** = p < 0.001\))} %feature on engagement}
\label{tab:profile_features}
\end{table}

\section{Discussion} \label{sec:discussion}

Our findings suggest that accounts that carry a higher level of information and emotional support help promote well-being. For instance, the uniform usage of second-person pronouns in the top categories suggests that communication, in general, is not about the self (the mental health expert) but rather inclusive of the community (as evident with words like `us'). This aligns with the greater use of second-person pronouns from therapist~\cite{valdes2010analysis}. This work also presents the first study on such accounts, and this characterization of accounts can be a basis for further research.

\pb{Professional Identity of Accounts.}
We show that accounts related to mental health can be classified based on their self-declared identities. Our results highlight that users' reactions to word categories differ based on the type of account that posts them. Thus, we argue that it is essential for individual accounts to adapt and focus their messages on their specific audiences.
A critical consideration for future studies is the role of self-description (a proxy to self-disclosure) regarding the audience's preference to interact with the account. Such disclosures from the professionals can impact both the users and the professionals~\cite{zur2009psychotherapist}. 

\pb{Staying relevant to topics.} Not all (semantic) topics receive equal levels of engagement. Topics that directly talk about mental health problems or use particular vocabulary, including causal indications, tend to get higher engagement. Interestingly our findings indicate that for mental health experts, including high information value (Information Support) in posts is more important than exhibiting empathy (Emotional Support).

\pb{Cross-Platform Influence.}
Instagram profiles that include links to other websites tend to get lower levels of engagement. Although other casual effects can potentially lead to such lowered engagement, this indicates that taking a more holistic multi-platform approach may be common among health influencers. Particularly, in our future work, we want to explore the relationships across social media platforms.

\pb{Limitations.} 
First, the number of likes as a measure of engagement is coarse and may mean different reactions from different people. Additionally, we cannot know \emph{when} or \emph{why} a user liked a post (as the API does not provide such data). To address this, we intend to inspect comments in our future work. Although fewer in number, these offer a richer insight that can be studied with respect to time.  
Second, we only use the textual captions of the posts. We posit that a combination of both images and text could reveal more about engagement. 
For example, it is possible that images may contain artifacts that impact engagement more than the caption. 
In the future, we plan to study these factors holistically. 
Finally, our curated set of 127 
accounts do not necessarily reflect all accounts well. Hence, we are cautious in over-generalizing our observations to other profiles or platforms.

\section{Conclusion} \label{sec:limitation}

We have explored the communication of 127 professional healthcare `influencers', 
as well as the engagement from their audiences. Our empirical analysis highlights this understudied group of mental health influencers. We find that these influencers offer various levels of informational supports (advice and knowledge) emotional supports (empathy and encouragement) to their followers. We have revealed links between the semantic context of words and the engagement rates of users.
Our findings highlight that the explicit discussion of negative emotion words (e.g., `\textit{anger}' and `\textit{trauma}') can raise the levels of user engagement. 
Our results suggest that exploring the communication cues between `influencers' and their followers could help us make better decisions about how to effectively support people with mental health needs. 

\section*{Acknowledgment}
This research has been supported by the 5GEAR project (Grant No. 318927) and the FIT project (Grant No. 325570) funded by the Academy of Finland.

\bibliographystyle{IEEEtran}
\bibliography{ref}

\begin{thebibliography}{10}
\providecommand{\url}[1]{#1}
\csname url@samestyle\endcsname
\providecommand{\newblock}{\relax}
\providecommand{\bibinfo}[2]{#2}
\providecommand{\BIBentrySTDinterwordspacing}{\spaceskip=0pt\relax}
\providecommand{\BIBentryALTinterwordstretchfactor}{4}
\providecommand{\BIBentryALTinterwordspacing}{\spaceskip=\fontdimen2\font plus
\BIBentryALTinterwordstretchfactor\fontdimen3\font minus
  \fontdimen4\font\relax}
\providecommand{\BIBforeignlanguage}[2]{{%
\expandafter\ifx\csname l@#1\endcsname\relax
\typeout{** WARNING: IEEEtran.bst: No hyphenation pattern has been}%
\typeout{** loaded for the language `#1'. Using the pattern for}%
\typeout{** the default language instead.}%
\else
\language=\csname l@#1\endcsname
\fi
#2}}
\providecommand{\BIBdecl}{\relax}
\BIBdecl

\bibitem{mythsfacts}
HSS, ``Mental health myths and facts,''
  \url{https://www.mentalhealth.gov/basics/mental-health-myths-facts}, 2017,
  (Accessed on 09/10/2022).

\bibitem{uk_survey_health_2014}
NHS, ``Survey of mental health,''
  \url{http://digital.nhs.uk/catalogue/PUB21748}, 2014, (Accessed on
  09/10/2022).

\bibitem{santarossa2018lancerhealth}
S.~Santarossa and S.~J. Woodruff, ``\# lancerhealth: Using twitter and
  instagram as a tool in a campus wide health promotion initiative,''
  \emph{Journal of Public Health Research}, vol.~7, no.~1, 2018.

\bibitem{de2014characterizing}
M.~De~Choudhury, S.~Counts, E.~J. Horvitz, and A.~Hoff, ``Characterizing and
  predicting postpartum depression from shared facebook data,'' in \emph{17th
  ACM CSCW}, 2014.

\bibitem{colbeck2008professional}
C.~L. Colbeck, ``Professional identity development theory and doctoral
  education,'' \emph{New Directions for Teaching and Learning}, vol. 2008, no.
  113, pp. 9--16, 2008.

\bibitem{caza2016construction}
B.~B. Caza and S.~Creary, ``The construction of professional identity,'' in
  \emph{Perspectives on contemporary professional work}.\hskip 1em plus 0.5em
  minus 0.4em\relax Edward Elgar Publishing, 2016.

\bibitem{larson1979rise}
M.~S. Larson and M.~S. Larson, \emph{The rise of professionalism: A
  sociological analysis}.\hskip 1em plus 0.5em minus 0.4em\relax Univ of
  California Press, 1979, vol. 233.

\bibitem{miller2017analyzing}
A.~M. Miller, ``Analyzing songs used for lyric analysis with mental health
  consumers using linguistic inquiry and word count (liwc) software,'' 2017.

\bibitem{valdes2010analysis}
N.~Vald{\'e}s, ``Analysis of patient and therapist linguistic styles during
  therapeutic conversation in change episodes, using linguistic inquiry and
  word count (liwc) 13,'' \emph{Faculty of Behavioural and Cultural Studies,
  Heidelberg University in Cooperation with the Pontificia Universidad
  Cat{\'o}lica de Chile}, p. 155, 2010.

\bibitem{de2017language}
M.~De~Choudhury and E.~Kiciman, ``The language of social support in social
  media and its effect on suicidal ideation risk,'' in \emph{Proceedings of the
  International AAAI ICWSM}, 2017.

\bibitem{peng2020exploring}
Z.~Peng, Q.~Guo, K.~W. Tsang, and X.~Ma, ``Exploring the effects of
  technological writing assistance for support providers in online mental
  health community,'' in \emph{Proceedings of the 2020 CHI Conference on Human
  Factors in Computing Systems}, 2020, pp. 1--15.

\bibitem{pennebaker2001linguistic}
J.~W. Pennebaker, M.~E. Francis, and R.~J. Booth, ``Linguistic inquiry and word
  count: Liwc 2001,'' \emph{Mahway: Lawrence Erlbaum Associates}, vol.~71, no.
  2001, p. 2001, 2001.

\bibitem{Kostoska2018real}
M.~Kostoska, M.~Simjanoska, B.~Koteska, and A.~M. Bogdanova, ``Real-time smart
  advisory health system,'' in \emph{Proceedings of the 8th International
  Conference on WIMS}, 2018.

\bibitem{kao2020user}
H.-T. Kao, S.~Yan, H.~Hosseinmardi, S.~Narayanan, K.~Lerman, and E.~Ferrara,
  ``User-based collaborative filtering mobile health system,'' \emph{Proc. ACM
  IMWUT}, vol.~4, no.~4, 2020.

\bibitem{de2013predicting}
M.~De~Choudhury, M.~Gamon, S.~Counts, and E.~Horvitz, ``Predicting depression
  via social media,'' in \emph{Seventh international AAAI ICWSM}, 2013.

\bibitem{cavlo2016computing}
R.~A. Calvo, K.~Dinakar, R.~Picard, and P.~Maes, ``Computing in mental
  health,'' ser. CHI EA '16.\hskip 1em plus 0.5em minus 0.4em\relax ACM, 2016.

\bibitem{doherty2008technologies}
G.~Doherty, J.~Sharry, M.~Bang, M.~Alca\~{n}iz, and R.~Ba\~{n}os,
  \emph{Technology in Mental Health}, 2008.

\bibitem{alqahtani2019usability}
F.~Alqahtani and R.~Orji, ``Usability issues in mental health applications,''
  ser. UMAP'19 Adjunct.\hskip 1em plus 0.5em minus 0.4em\relax ACM, 2019.

\bibitem{haq_2022_govt}
E.-U. Haq, T.~Braud, L.~H. Lee, R.~H. Mogavi, H.~Zhang, and P.~Hui, ``Tips,
  tidings, and tech: Governmental communication on facebook during the covid-19
  pandemic,'' in \emph{DG.O 2022: The 23rd Annual International Conference on
  Digital Government Research}, 2022.

\bibitem{villani_does_2012}
D.~Villani and G.~Riva, ``Does {Interactive} {Media} {Enhance} the {Management}
  of {Stress}? {Suggestions} from a {Controlled} {Study},''
  \emph{Cyberpsychology, Behavior, and Social Networking}, 2012.

\bibitem{krause2001social}
N.~Krause, ``Social support.'' 2001.

\bibitem{kaplan1977social}
B.~H. Kaplan, J.~C. Cassel, and S.~Gore, ``Social support and health,''
  \emph{Medical care}, vol.~15, no.~5, pp. 47--58, 1977.

\bibitem{glanz2008health}
K.~Glanz, B.~K. Rimer, and K.~Viswanath, \emph{Health behavior and health
  education: theory, research, and practice}.\hskip 1em plus 0.5em minus
  0.4em\relax John Wiley \& Sons, 2008.

\bibitem{nick2018online}
E.~A. Nick, D.~A. Cole, S.-J. Cho, D.~K. Smith, T.~G. Carter, and R.~L.
  Zelkowitz, ``The online social support scale: Measure development and
  validation.'' \emph{Psychological assessment}, vol.~30, no.~9, p. 1127, 2018.

\bibitem{harandi2017correlation}
T.~F. Harandi, M.~M. Taghinasab, and T.~D. Nayeri, ``The correlation of social
  support with mental health: A meta-analysis,'' \emph{Electronic physician},
  vol.~9, no.~9, p. 5212, 2017.

\bibitem{kandappu2018feasibility}
T.~Kandappu, A.~Misra, D.~Koh, R.~D. Tandriansyah, and N.~Jaiman, ``A
  feasibility study on crowdsourcing to monitor municipal resources in smart
  cities,'' in \emph{Companion Proceedings of the The Web Conference 2018},
  2018, pp. 919--925.

\bibitem{haq_screenshots_2022}
E.-U. Haq, T.~Braud, Y.-P. Yau, L.-H. Lee, F.~B. Keller, and P.~Hui,
  ``Screenshots, symbols, and personal thoughts: The role of instagram for
  social activism,'' in \emph{Proceedings of the ACM Web Conference
  2022}.\hskip 1em plus 0.5em minus 0.4em\relax New York, NY, USA: Association
  for Computing Machinery, 2022.

\bibitem{rossi2018venice}
L.~Rossi, E.~Boscaro, and A.~Torsello, ``Venice through the lens of instagram:
  A visual narrative of tourism in venice,'' in \emph{Companion Proceedings of
  the The WebConf}, 2018.

\bibitem{zhang2019language}
Y.~Zhang, ``Language in our time: An empirical analysis of hashtags,'' in
  \emph{The World Wide Web Conference}, 2019.

\bibitem{zarei2020characterising}
K.~Zarei, D.~Ibosiola, R.~Farahbakhsh, Z.~Gilani, K.~Garimella, N.~Crespi, and
  G.~Tyson, ``Characterising and detecting sponsored influencer posts on
  instagram,'' in \emph{IEEE/ACM ASONAM}, 2020.

\bibitem{ferwerda2015predicting}
B.~Ferwerda, M.~Schedl, and M.~Tkalcic, ``Predicting personality traits with
  instagram pictures,'' in \emph{Proceedings of the 3rd Workshop on Emotions
  and Personality in Personalized Systems 2015}, 2015, pp. 7--10.

\bibitem{baker2019qualitative}
N.~Baker, G.~Ferszt, and J.~G. Breines, ``A qualitative study exploring female
  college students' instagram use and body image,'' \emph{Cyberpsychology,
  behavior, and social networking}, vol.~22, no.~4, pp. 277--282, 2019.

\bibitem{ahadzadeh2017self}
A.~S. Ahadzadeh, S.~P. Sharif, and F.~S. Ong, ``Self-schema and
  self-discrepancy mediate the influence of instagram usage on body image
  satisfaction among youth,'' \emph{Computers in Human Behavior}, vol.~68, pp.
  8--16, 2017.

\bibitem{santos2020instagram}
A.~Santos and M.~Figueras, ``Instagram and gender inequalities: The discourse
  of young women regarding social networks,'' in \emph{Eighth International
  Conference on Technological Ecosystems for Enhancing Multiculturality}, 2020,
  pp. 577--581.

\bibitem{pinto2020instagram}
P.~A. Pinto, M.~J.~L. Antunes, and A.~M.~P. Almeida, ``Instagram as a
  communication tool in public health: a systematic review,'' in \emph{2020
  15th Iberian Conference on Information Systems and Technologies
  (CISTI)}.\hskip 1em plus 0.5em minus 0.4em\relax IEEE, 2020, pp. 1--6.

\bibitem{griffith2021mentalhealthart}
F.~J. Griffith, C.~H. Stein, J.~E. Hoag, and K.~N. Gay, ``\# mentalhealthart:
  How instagram artists promote mental health awareness online,'' \emph{Public
  Health}, vol. 194, pp. 67--74, 2021.

\bibitem{pinto2021public}
P.~A. Pinto, M.~J.~L. Antunes, and A.~M.~P. Almeida, ``Public health on
  instagram: an analysis of health promotion strategies of portugal and
  brazil,'' \emph{Proc. Computer Science}, 2021.

\bibitem{andalibi2017sensitive}
N.~Andalibi, P.~Ozturk, and A.~Forte, ``Sensitive self-disclosures, responses,
  and social support on instagram: the case of\# depression,'' in \emph{2017
  ACM conference on CSCW}, 2017.

\bibitem{centola2011experimental}
D.~Centola, ``An experimental study of homophily in the adoption of health
  behavior,'' \emph{Science}, vol. 334, no. 6060, 2011.

\bibitem{bojanowski2016enriching}
P.~Bojanowski, E.~Grave, A.~Joulin, and T.~Mikolov, ``Enriching word vectors
  with subword information,'' \emph{arXiv preprint arXiv:1607.04606}.

\bibitem{pennebaker2015development}
J.~W. Pennebaker, R.~L. Boyd, K.~Jordan, and K.~Blackburn, ``The development
  and psychometric properties of liwc2015,'' Tech. Rep., 2015.

\bibitem{wang2012stay}
Y.-C. Wang, R.~Kraut, and J.~M. Levine, ``To stay or leave? the relationship of
  emotional and informational support to commitment in online health support
  groups,'' in \emph{Proceedings of the ACM 2012 CSCW}, 2012, pp. 833--842.

\bibitem{bakhshi2014faces}
S.~Bakhshi, D.~A. Shamma, and E.~Gilbert, ``Faces engage us: Photos with faces
  attract more likes and comments on instagram,'' in \emph{ACM CHI}, 2014.

\bibitem{zur2009psychotherapist}
O.~Zur, M.~H. Williams, K.~Lehavot, and S.~Knapp, ``Psychotherapist
  self-disclosure and transparency in the internet age.'' \emph{Professional
  Psychology: Research and Practice}, 2009.

\end{thebibliography}

\end{document}